\documentclass[aps,pra,showpacs,preprint]{revtex4}
\usepackage{graphicx}
\usepackage{amsmath,amssymb,amsfonts}
\usepackage{dcolumn}
\usepackage{epsfig}

\begin{document}

\title{Application of the $J$-matrix Method to Faddeev-Merkuriev equation:
beyond pseudostates.}
\author{ S. A. Zaytsev}
\email[E-mail: ]{zaytsev@fizika.khstu.ru} \affiliation{Pacific
National University, Khabarovsk, 680035, Russia}
\author{ V. A. Knyr}
\affiliation{Pacific National University, Khabarovsk, 680035,
Russia}
\author{Yu. V. Popov}
\affiliation{Nuclear Physics Institute, Moscow State University,
119992, Moscow}
\author{ A. Lahmam-Bennani}
\affiliation{Laboratoire des Collisions Atomiques et Moleculaires
(UMR 8625) and Federation Lumiere Matiere(FR 2764), Bat. 351,
Universit\'e de Paris-Sud XI, F91405 Orsay Cedex,France}

\begin{abstract}
A version of the $J$-matrix method for solving numerically the
three-body Faddeev-Merkuriev differential equations is proposed.
This version allows to take into account the full spectrum of the
two-body Coulomb subsystem. As a result, a discrete analog of the
Lippmann-Schwinger equation is obtained which allows to interpret
correctly the three-body wave function in two-body domains. The
scheme is applied to calculations of the fully resolved absolute
differential cross sections for the He$(e,2e)$He$^+$ and
He$(e,3e)$He$^{++}$ reactions at small energy and momentum
transfers. The results are in good agreement with the experiment
both in shape and in absolute value.
\end{abstract}
\maketitle

\section{Introduction}

The few-body Coulomb problem is one of the most fundamental
problems in physics and being yet to be exactly solved. In recent
years, it has attracted numerous theoretical studies which involve
different approaches and/or approximations. Among these is the
$J$-matrix method widely used in the quantum scattering theory. It
was first proposed in atomic physics~\cite{JA} and later on
(independently) in nuclear physics~\cite{FO,SN,RSJ}. Within this
approach the full Hamiltonian of the atomic system is split into
two parts $H=H_c+V$. The operator $H_c$ determines the discrete
basis of square-integrable functions which are used for expansion
of the many-body wave function. In atomic physics, a Laguerre
basis is the most widely used because it provides the
three-diagonal representation of a radial part of the operator
$H_c$, and this infinite trinomial recurrence can be solved
analytically.  At the same time, the short-range interaction $V$
is approached by its projection $V^N$ on a subspace of $N$ basic
functions. Hence, the $J$-matrix method supplies the exact
solution of the scattering problem on the model potential $V^N$.
The $J$-matrix approach is proved to be an efficient and rather
accurate numerical method. Formally and from the viewpoint of a
numerical realization, the $J$-matrix method is similar to the
R-matrix scattering theory (for an overview, see, for
example,~\cite{Rmtx}). It is also equivalent to the method of the
so-called Coulomb-Sturm separable expansion (see, for
instance,~\cite{Papp} and references therein).

The most important problem in a few-body Coulomb scattering theory
is the description of the continuum-state wave function in terms
of square-integrable functions. The method of pseudostates and its
recent avatar, the convergent close-coupling method
(CCC)~\cite{BS}, replace the continuous energy spectrum of a
selected two-body subsystem by a finite number of positive energy
lines. However, we present here a new scheme which allows to take
into account both the summation and integration over respectively
the bound and continuum states of the two-body subsystem.

The general idea guiding the $J$-matrix method was first
formulated in~\cite{KS} for a system of three nuclear particles
interacting with short-range potentials using an oscillator basis.
Recently, on a base of the results of Papp et al.~\cite{PHHKY} it
was extended for the case of long-range Coulomb
potentials~\cite{ZKP}. We solve the Faddeev-Merkuriev differential
equations which allow to formulate the boundary conditions for a
component $\psi_{\alpha}$ of the full wave function
$\Psi=\sum_{\alpha} \psi_{\alpha}$ in terms of the fixed number of
Jacobi coordinates ${\bf x}_{\alpha},{\bf y}_{\alpha}$. We use the
fact \cite{MFB} that in the two-body domain $\Omega_{\alpha}$,
where the mutual distance ${\bf x}_{\alpha}$ between particles
$\beta$ and $\gamma$ is much smaller than the distance ${\bf
y}_{\alpha}$ between their center-of-mass and particle $\alpha$,
the total asymptotic Hamiltonian $H_{as}$ looks like a direct sum
of uncoupled Hamiltonians $h_{x_{\alpha}}$ and $h_{y_{\alpha}}$.
As a result, the full three-body Green's function $G=(E-H+i0)$ is
connected with the asymptotic operator
$G_{\alpha}=(E-h_{x_{\alpha}}-h_{y_{\alpha}}+i0)$ in the domain
$\Omega_{\alpha}$. In turn, the Green's function $G_{\alpha}$ is
presented as an overlapping integral of the operators
$G_{x_{\alpha}}$ and $G_{y_{\alpha}}$.

Furthermore, following Ref.~\cite{ZKP} the component
$\psi_{\alpha}$ is presented as a decomposition on the
eigenfunctions of the two-body subsystem $(\beta\gamma)$ with the
total charge $Z_{\alpha}$. Such a choice allows to factor out the
long-range part of the interaction and, consequently, to describe
correctly the asymptotic behavior of the function $\Psi$ only in
the two-body domain $\Omega_{\alpha}$. In this way, we arrive at a
discrete analog of the Lippmann-Schwinger equation for the
component $\psi_{\alpha}$ in $\Omega_{\alpha}$. This is the
general scheme of the method, and the details are presented below.

The present paper is organized as follows. A detailed theoretical
formulation of the above general scheme is presented in Section
II. In Section III, the efficiency of the presented numerical
scheme is demonstrated in a light of comparison with experimental
data. We calculate for a helium target the $(e,2e)$ triple
differential cross section (TDCS) for single ionization
accompanied with simultaneous excitation, and the $(e,3e)$
five-fold differential cross section (5DCS) for double ionization
under fast electron impact. Previous calculations of the
three-body wave function of the system $(e,e,${He}$^{++})$ were
carried out within the method of pseudostates~\cite{KBLDT,KNP} as
well as using other methods, such as the so-called
3C~\cite{JM,AMC}, approximate 6C~\cite{JM,6C}, and C4FS
\cite{C4FS}. Most of theoretical results exhibited a marked
disagreement with the experimental data on absolute
scale~\cite{KBLDT,KNP}. In this work we show that the proposed
method is able to describe the same experiments without any
scaling factors.

\section{Theory}

The Hamiltonian of a three-body system has the form
\begin{equation}
H=H_0 +\sum \limits _{\alpha=1}^3 V^C_{\alpha}(x_{\alpha}),
 \label{H3}
\end{equation}
where $H_0$ is the kinetic energy operator
\begin{equation}
H_0= -\triangle_{x_{\alpha}}-\triangle_{y_{\alpha}},\label{KE}
\end{equation}
and
\begin{equation}
V^C_{\alpha}(x_{\alpha})=\frac{Z_{\alpha}}{x_{\alpha}}. \label{VC}
\end{equation}
The couple (${\bf x}_{\alpha}$, ${\bf y}_{\alpha}$) stands for the
set of Jacobi coordinates \cite{MFB}
\begin{equation}
 \begin{array}{c}
{\bf x}_{\alpha}=\tau_{\alpha}({\bf r}_{\beta}-{\bf r}_{\gamma}),\\[3mm]
{\bf y}_{\alpha}= \mu_{\alpha}\left({\bf r}_{\alpha}-
\frac{m_{\beta}{\bf r}_{\beta}+m_{\gamma}{\bf
r}_{\gamma}}{m_{\beta}+m_{\gamma}}\right) \ ,\\
 \end{array}
\label{Jc}
\end{equation}
where $m_i$ are the particle masses and
\begin{equation}
\tau_{\alpha}=\sqrt{2\frac{m_{\beta}m_{\gamma}}{m_{\beta}+m_{\gamma}}},
\quad
\mu_{\alpha}=\sqrt{2m_{\alpha}\left(1-\frac{m_{\alpha}}{m_1+m_2+m_3}
\right)}.
\end{equation}

The interaction $V_{\alpha}$ can be decomposed into the short- and
long-range parts ($V_{\alpha}^{(s)}$ and $V_{\alpha}^{(l)}$,
respectively) \cite{MFB}
\begin{equation}
V_{\alpha}^{(s)}(x_{\alpha}, \, y_{\alpha})=
V_{\alpha}(x_{\alpha}) \zeta_{\alpha}(x_{\alpha}, \, y_{\alpha}),
\qquad  V_{\alpha}^{(l)}(x_{\alpha}, \, y_{\alpha})=
V_{\alpha}(x_{\alpha})[1- \zeta_{\alpha}(x_{\alpha}, \,
y_{\alpha})]\label{SLD}
\end{equation}
with the "separation" function of the form \cite{KWH}
\begin{equation}
\zeta(x_{\alpha},y_{\alpha}) =
2/\left\{1+\exp[(x_{\alpha}/x_0)^{\nu}/(1+y_{\alpha}/y_0)]
\right\} \label{zeta}
\end{equation}
where $\nu >2$. Thus, the function $V_{\alpha}^{(s)}$ decreases
rather rapidly in the "true" three-body asymptotic domain
$\Omega_0$ and coincides with the initial potential in the
two-body asymptotic domain $\Omega_{\alpha}$ ($x_{\alpha} \ll
y_{\alpha}$). In the general case of different particles, the
total wave function is represented as the sum of three components
$\psi_{\alpha}$ satisfying the set of equations \cite{MFB}
\begin{equation}
\left[H_0+V_{\alpha}(x_{\alpha})+V_{\beta}^{(l)}(x_\beta)
+V_{\gamma}^{(l)}(x_{\gamma})-E\right] \psi_{\alpha} =
-V_{\alpha}^{(s)}\left(\psi_{\beta}+\psi_{\gamma} \right).
\label{FMEq}
\end{equation}
For the system $(e,e,{He}^{++})$ particles 1 and 2 (electrons) are
identical, and the solution reduces to the sum of two components
$\psi_1$ and $\psi_2$ (see, for instance, \cite{KWH}). They are
related to each other as follows: $\psi_2=gP_{12}\psi_1$ ($g=+1$
and $g=-1$ for, respectively, singlet and triplet spin states,
$P_{12}$ is the permutation operator). Taking into account the
spatial symmetry of the total wave function, Eq.~(\ref{FMEq}) can
be reduced to a single equation \cite{KWH}
\begin{equation}
[H_0+V_1(x_1)+V_3(x_3)+V_2^{(l)}(x_2)-E]\psi_1({\bf x}_1, \, {\bf
y}_1) = -gV_1^{(s)}P_{12}\psi_1({\bf x}_1, \, {\bf y}_1)
\label{FME}
\end{equation}
for the component $\psi_1$.

The wave function $\Psi^{(-)}$ of the system $(e,e,${He}$^{++})$
with all three particles being free can be represented by the
following decomposition
\begin{equation}\label{FWF}
  \Psi^{(-)}= \frac{2}{\pi}\frac{1}{p_0\,k_0}\,\frac{(1+gP_{12})}{\sqrt{2}}\,\sum
  \limits_{\substack{
  L\,\ell_0\,\lambda_0 \\ m_0\,\mu_0 }}
  (\ell_0\,m_0\,\lambda_0\,\mu_0|L\,M)
  {i}^{\ell_0+\lambda_0}\,e^{-{i}(\sigma_{\ell_0}+\sigma_{\lambda_0})}
  Y_{\ell_0\,m_0}^*(\widehat{k}_0)Y_{\lambda_0\,\mu_0}^*(\widehat{p}_0)
  \,\psi^{L\,M}_{\ell_0\,\lambda_0}.\\
\end{equation}
The corresponding decomposition of the function $\Psi^{(-)}_{n_0
\ell_0 m}$ of the system $(e,${He}$^+$) with two bound particles
is given by
\begin{equation}
 \label{FWFp}
  \Psi^{(-)}_{n_0\,\ell_0\,m}= \sqrt{\frac{2}{\pi}}\frac{1}{p_0}\,\frac{(1+gP_{12})}{\sqrt{2}}\,
  \sum  \limits_{\  L\,\lambda_0\,\mu_0}
  (\ell_0\,m\,\lambda_0\,\mu_0|L\,M)\,{i}^{\lambda_0}\,e^{-{
  i}\sigma_{\lambda_0}}\,Y_{\lambda_0\,\mu_0}^*(\widehat{p}_0)
  \,\psi^{L\,M}_{\ell_0\,\lambda_0},
\end{equation}
where the quantum numbers $(n_0\,\ell_0\,m)$ describe the state of
the He$^+$ ion.

The spatial part $\psi^{L\,M}_{\ell_0\,\lambda_0}$ of the
component $\psi_1$ in (\ref{FWF}) and (\ref{FWFp}) can be
presented in the form of a bispherical expansion
\begin{equation}\label{Space}
\psi^{L\,M}_{\ell_0\,\lambda_0}({\bf x},\,{\bf y})=\sum \limits
_{\ell\;
\lambda}\frac{\psi^L_{\ell_0\,\lambda_0,\;\ell\,\lambda}(x,\,y)}{xy}\mathcal{Y}
_{\ell\,\lambda}^{LM}(\widehat{x},\,\widehat{y}), \quad
 {\bf x}\equiv {\bf x}_1, \quad {\bf y}\equiv {\bf y}_1,
\end{equation}
\begin{equation}
\mathcal{Y} _{\ell\,\lambda}^{LM}(\widehat{x},\,\widehat{y})=
\sum \limits _{m\,
\mu}(\ell\,m\,\lambda\,\mu|L\,M)\,Y_{\ell\,m}(\widehat{x})\,
Y_{\lambda\,\mu}(\widehat{y}). \ \label{AF}
\end{equation}
In turn, and in accord with \cite{KS}, we look for the radial
functions $\psi^L_{\ell_0\,\lambda_0,\;\ell\,\lambda}(x,\,y)$ in
the form of an expansion
\begin{equation}\label{Rad}
\psi^L_{\ell_0\,\lambda_0,\;\ell\,\lambda}(x,\,y)=\sum \limits
_{\nu}\int dk a_{\nu}^{\ell\,\lambda}(k)
\varphi_{k,\ell}(x)\phi_{\nu}^{\lambda}(y)
\end{equation}
in the eigenfunctions $\varphi_{k,\ell}(x)$ of the Hamiltonian
\begin{equation}
h_x=-\triangle_x+\frac{Z_1}{x} \label{hx}
\end{equation}
describing the subsystem $(2,\,3)$. Note that here we pave the way
for nondemocratic successive ejection of electrons from the atom.

In (\ref{Rad}) and hereafter $\int dk$ means the summation over
the discrete states and integration over the continuous states of
the subsystem $(2,\,3)$, i.e.
\begin{equation}\label{Idk}
  \int dk a_{\nu}^{\ell\,\lambda}(k)\varphi_{k,\ell}(x)=\sum \limits _j
\varphi_{\ell}^{(j)}(x)\,a_{\nu}^{\ell\, \lambda}({i}\kappa_j)+
 \frac{2}{\pi}\int \limits _{0}^{\infty}
 dk \, \varphi_{\ell}(k,\,x)\,a_{\nu}^{\ell\, \lambda}(k),
\end{equation}
with $\varphi^{(j)}_{\ell}$ and $\varphi_{\ell}(k,\,x)$ being the
corresponding eigenfunctions of the Hamiltonian (\ref{hx}) [see
Eqs.~(\ref{ABndf}) and (\ref{ARegSol}) of Appendix]. The Laguerre
basis functions $\phi_{\nu}^{\lambda}$ \cite{JA} are used in
(\ref{Rad})
\begin{equation}
\phi_{\nu}^{\lambda}(r)=\left[(\nu+1)_{(2\lambda+1)}
\right]^{-1/2}(2ur)^{\lambda+1}e^{-ur}L_{\nu}^{2\lambda+1}(2ur),
\label{BF}
\end{equation}
with $u$ being the basis parameter which suitable choice affects
the rate of convergence of the numerical results.

The functions $\varphi^{(j)}_{\ell}$ and $\varphi_{\ell}(k,\,x)$
can be also decomposed on the basis functions $\phi_{n}^{\ell}$
(\ref{BF}). We use the analytical expressions for the coefficients
$\mathcal{S}_{n\,\ell}^{(j)}$ and $\mathcal{S}_{n \ell}(k)$ of
such a decomposition [see Eqs.~(\ref{ASB}) and (\ref{ASsol}) of
Appendix]. Thus, the function $\psi^{L\,M}_{\ell_0\,\lambda_0}$
reads
\begin{equation}
\psi^{L\,M}_{\ell_0\,\lambda_0} = \sum \limits _{\ell, \,
\lambda, \, n, \, \nu}
C_{n\,\nu}^{L\,(\ell\,\lambda)}(E)\,\left|n\,\ell\, \nu \,
\lambda; \; LM \right>, \label{BE}
\end{equation}
\begin{equation}
\left|n\,\ell\, \nu \, \lambda; \; LM \right> =
\frac{\phi_{n}^{\ell}(x)\,\phi_{\nu}^{\lambda}(y)}{xy}\,
\mathcal{Y}^{LM}_{\ell\, \lambda}(\widehat{x}, \,\widehat{y}).
\label{BBF}
\end{equation}
The coefficients $C_{n\,\nu}^{L\,(\ell\,\lambda)}$ are of the form
\begin{equation}
C_{n\,\nu}^{L\,(\ell\,\lambda)}(E) = \int dk \, \mathcal{S}_{n
\ell}(k)\,a_{\nu}^{\ell\, \lambda}(k). \label{EC}
\end{equation}

The set of algebraic equations for the coefficients
$a_{\nu}^{\ell\, \lambda}(k)$ was obtained in the papers
\cite{KS,ZKP}. As a result, Eq. (\ref{FME}) transforms to the form
\begin{equation} \mathcal{J}_{\nu\,\nu''}^{\lambda}(p) \,
a_{\nu''}^{\ell\,\lambda}(k)=-\sum \limits_{\substack{n, \, n',\,
\nu' \\ \ell', \, \lambda'}}\mathcal{S}_{n\,\ell}(k)\,
V^{L(\ell\,\lambda)(\ell'\,\lambda')}_{n\,\nu,\;
n'\,\nu'}\,C_{n'\,\nu'}^{L\,(\ell'\,\lambda')}(E),\quad k^2+p^2=E.
\label{Ja}
\end{equation}
Here $\mathcal{J}_{\nu\,\nu'}^{\lambda}$ are the elements of the
$J$-matrix  [see Eq.~(\ref{AJMT}) of Appendix] corresponding to the
operator $(h_y-p^2)$ ($h_y=-\triangle_y+{Z_{11}}/{y}$). The
potential ${Z_{11}}/{y}$ describes the Coulomb interaction of the
particle 1 with the center-of-mass of the subsystem $(2, \, 3)$, and
values $V^{L(\ell\,\lambda)(\ell'\,\lambda')}_{n\,\nu,\; n'\,\nu'}$
denote the matrix elements of the operator
\begin{equation}
V({\bf x}, {\bf y})=
V_3(x_3)+V_2^{(l)}(x_2)-\frac{Z_{11}}{y}+gV_1^{(s)}P_{12},\label{VSH}
\end{equation}
i.e.
\begin{equation}
V^{L(\ell\,\lambda)(\ell'\,\lambda')}_{n\,\nu,\; n'\,\nu'}= \left<
n\,\ell\, \nu \, \lambda; \; LM \right| V({\bf x}, \, {\bf y})
\left| n'\,\ell'\, \nu' \, \lambda'; \; LM\right>. \label{VSHME}
\end{equation}
Within the two-body domain where $x<<y\rightarrow\infty$, the
potential $V({\bf x}, {\bf y})$ is a short-range one. This
circumstance allows to take into account only a finite number of
terms in the r.h.s. of Eq.~(\ref{Ja}). This means that
$V^{L(\ell\,\lambda)(\ell'\,\lambda')}_{n\,\nu,\; n'\,\nu'}=0$ if
at least one of the indexes $\{ n,\,\nu,\, n',\,\nu' \}$ extends
to some rather large number $N$. Thus, if $\nu \ge N$ then the
r.h.s. of eq. (\ref{Ja}) turns to zero, and the coefficients
$a_{\nu}^{\ell\,\lambda}$ satisfy the "free" equation
\begin{equation}
\mathcal{J}_{\nu\,\nu-1}^{\lambda}(p)\,a_{\nu-1}^{\ell\,\lambda}(k)+
\mathcal{J}_{\nu\,\nu}^{\lambda}(p)\,a_{\nu}^{\ell\,\lambda}(k)+
\mathcal{J}_{\nu\,\nu+1}^{\lambda}(p)\,a_{\nu+1}^{\ell\,\lambda}(k)=0.
\label{TRR}
\end{equation}

Now we use the expression \cite{Heller}
\begin{equation}
\mathcal{G}_{\nu\, \nu'}^{\lambda(\pm)}(p)
=-\frac{1}{p}\,\mathcal{S}_{\nu_{<}\,\lambda}(p)\,
\mathcal{C}_{\nu_{>}\,\lambda}^{(\pm)}(p), \quad
\nu_{<}=\min\left\{\nu, \, \nu'\right\}, \;
\nu_{>}=\max\left\{\nu, \, \nu'\right\},\\
\label{GFME}
\end{equation}
for the matrix elements $\left[\mathcal{G}^{\lambda (\pm)}(p)
\right]$ of the radial Green's function. The functions
$\mathcal{C}_{\nu\,\lambda}^{(\pm)}(p)$ are determined by
Eq.~(\ref{ASCp}) of Appendix. Since the infinite matrix
$\left[\mathcal{G}^{\lambda (\pm)}(p) \right]$ formally is the
inverse matrix to the infinite $J$-matrix
$\left[\mathcal{J}^{\lambda}(p)\right]$ \cite{Heller}, i.e.
\begin{equation}
-\left[\mathcal{G}^{\lambda (\pm)}(p)\right]\,
\left[\mathcal{J}^{\lambda}(p)\right] = 1,\label{GJ}
\end{equation}
the set of equations (\ref{Ja}) can be rewritten in the following
form
\begin{equation}
 \begin{array}{l}
a_{\nu}^{\ell\,\lambda}(k)= \mathcal{S}_{\nu\,
\lambda_0}(p_0)\delta_{(\ell\,\lambda)\,(\ell_0\,\lambda_0)}\delta(k-k_0)+\\[3mm]
\qquad \qquad +\sum \limits_{\substack{n', \, \nu', \, n'', \, \nu''
\\ \ell'', \, \lambda''}}\mathcal{G}_{\nu\,
\nu'}^{\lambda(-)}(p)\mathcal{S}_{n'\,\ell}(k)\,
V^{L(\ell\,\lambda)(\ell''\,\lambda'')}_{n'\,\nu',\;
n''\,\nu''}\,C_{n''\,\nu''}^{L\,(\ell''\,\lambda'')}(E), \quad \nu=0, \, 1, \ldots.\\
\end{array} \label{LSa}
\end{equation}
Inserting  (\ref{LSa}) into (\ref{EC}) and taking into account the
boundary conditions for the functions $\Psi^{(-)}$ (\ref{FWF}) and
$\Psi^{(-)}_{n_0\,\ell_0\,m}$ (\ref{FWFp}), we obtain the set of
equations for the expansion coefficients
$C_{n\,\nu}^{L\,(\ell\,\lambda)}$ in (\ref{BE}). In the case,
where all particles are asymptotically free, it takes the form
\begin{equation}
 \begin{array}{c}
C_{n\,\nu}^{L\,(\ell\,\lambda)}(E)=
\delta_{(\ell\,\lambda)\,(\ell_0\,\lambda_0)}\,\mathcal{S}_{n
\ell_0}(k_0)\, \mathcal{S}_{\nu \lambda_0}(p_0)+\qquad \qquad
\qquad \qquad \qquad \qquad \qquad \qquad \\[5mm]
\qquad \qquad +\sum \limits _{\substack{n', \, \nu', \, n'', \,
\nu'' \\ \ell'', \, \lambda''}}\left[  \int
dk\,\mathcal{S}_{n\ell}(k)\, \mathcal{S}_{n'\,\ell}(k)\,
\mathcal{G}^{\lambda (-)}_{\nu\, \nu'}(p)
\right]\,V^{L(\ell\,\lambda)(\ell''\,\lambda'')}_{n'\,\nu',\;
n''\,\nu''}\,C_{n''\,\nu''}^{L\,(\ell''\,\lambda'')}(E)\\
 \end{array}
\label{CE}
\end{equation}
In the case of two asymptotically bound particles the set of
equations takes the form
\begin{equation}
 \begin{array}{c}
C_{n\,\nu}^{L\,(\ell\,\lambda)}(E)=
\delta_{(\ell\,\lambda)\,(\ell_0\,\lambda_0)}\,\mathcal{S}_{n
\ell_0}^{(n_0)}\, \mathcal{S}_{\nu \lambda_0}(p_0)+\qquad \qquad
\qquad \qquad \qquad \qquad \qquad \qquad \\[5mm]
\qquad \qquad +\sum \limits _{\substack{n', \, \nu', \, n'' \, \nu''
\\ \ell'', \, \lambda''}}\left[  \int dk\,\mathcal{S}_{n\ell}(k)\,
\mathcal{S}_{n'\,\ell}(k)\, \mathcal{G}^{\lambda (-)}_{\nu\,
\nu'}(p)
\right]\,V^{L(\ell\,\lambda)(\ell''\,\lambda'')}_{n'\,\nu',\;
n''\,\nu''}\,C_{n''\,\nu''}^{L\,(\ell''\,\lambda'')}(E)\\
 \end{array}
\label{CEp}
\end{equation}

The set of equations (\ref{CE}) and (\ref{CEp}) is in fact the
discrete analog of an integral equation of the Lippmann-Schwinger
type, which was obtained in \cite{PHHKY}. It is easy to see that
the integral in square brackets of (\ref{CE}) coincides with the
matrix elements $G_{n\,n';\;\nu\,\nu'}^{\ell\,\lambda\,(-)}$ of
the asymptotic three-body Green's function $\widetilde{G}^{(-)}$
in the two-body domain. It has the form of a contour overlap
integral of the two-body Green's functions $G^{(-)}_x$ and
$G^{(-)}_y$ \cite{MFB}
\begin{equation}\label{G3}
  \widetilde{G}^{(-)}(E)=
  \frac{1}{2\pi {i}} \oint
  \limits _{\mathcal{C}}d\mathcal{E}\,G_x^{(-)}(\mathcal{E})
G_y^{(-)}(E-\mathcal{E}).
\end{equation}
Here the contour $\mathcal{C}$ surrounds the spectrum of the
operator $h_x$ in a anticlockwise direction.

The integrals in the r.h.s. of Eq.~(\ref{CE}) can be calculated
directly. Note that the integrand has the following poles at the
points  $k_j=\sqrt{E+\eta^2_j}$
($\eta_j={-Z_{11}}/[{2(\lambda+j+1)}]$) which correspond to the
discrete spectrum of the Hamiltonian $h_y$ and have a densening
point at $k_{\infty}=\sqrt{E}$. The presence of the matrix element
of the radial Green's function $\mathcal{G}^{\lambda (-)}_{\nu\,
\nu'}(p)$ in the integrand means that the poles have to be rounded
in a clockwise direction (Fig.~1).

Let us note that the poles can be allocated more uniformly along the
contour of integration if one carries out the transformation of
variable.  For example, if one puts $q={-Z_{11}}/(2\sqrt{k^2-E})$
then the poles are disposed at the points $q_j=\lambda+1+j$. Thus,
the integral under consideration can be performed as an infinite sum
of integrals $I_j$ along segments $L_j$, in which the poles are
localized, and integral $I_e$ with a smooth integrand, which is
calculated along the segment $L_e$ disposing to the right of the
segment $L_0$. Using the Sokhotsky formula we obtain
\begin{equation}\label{Ij}
 \begin{array}{c}
  I_j = \frac{2}{\pi}\, \lefteqn{-}{\int \limits _{L_j}}
dk \mathcal{S}_{n\ell}(k)\, \mathcal{S}_{n'\,\ell}(k)\,
\mathcal{G}^{\lambda (-)}_{\nu\, \nu'}(p)
 -2{i}\, \mathcal{S}_{n\ell}(k_j)\mathcal{S}_{n'\,\ell}(k_j)
 \mathop{Res}\limits_{k=k_j}\mathcal{G}^{\lambda (-)}_{\nu\,
\nu'}(p)=\\[3mm]
\qquad \qquad \qquad =\frac{2}{\pi}\, \lefteqn{-}{\int \limits
_{L_j}} dk \mathcal{S}_{n\ell}(k)\, \mathcal{S}_{n'\,\ell}(k)\,
\mathcal{G}^{\lambda (-)}_{\nu\, \nu'}(p)
 +\frac{{i}}{k_j}
 \mathcal{S}_{n\ell}(k_j)\mathcal{S}_{n'\,\ell}(k_j)\,
 \mathcal{S}_{\nu\,\lambda}^{(j)}\,\mathcal{S}_{\nu'\,\lambda}^{(j)}.\\
 \end{array}
\end{equation}

For $E<0$ the matrix elements
$G_{n\,n';\;\nu\,\nu'}^{\ell\,\lambda\,(-)}$ can be calculated
without difficulties because the number of poles is finite and
(usually) not numerous.  For $E>0$ we must calculate an infinite
number of integrals in (\ref{Ij}). In this particular case we
calculate the overlap integral (\ref{G3}) being written now for
the matrix elements
\begin{equation}\label{GC}
  G_{n\,n';\;\nu\,\nu'}^{\ell\,\lambda\,(-)}(E)=
  \frac{1}{2\pi {i}} \int
  \limits _{\mathcal{C}}d\mathcal{E}\,\mathcal{G}^{\ell\,(-)}_{n\,
n'}(\sqrt{\mathcal{E}_0+\mathcal{E}}) \mathcal{G}^{\lambda
(-)}_{\nu\, \nu'}(\sqrt{\mathcal{E}_0-\mathcal{E}}),
\end{equation}
with $\mathcal{E}_0 = \frac12 E$, and use the method of Shakeshaft
\cite{RS}. The poles and cuts of the integrand in (\ref{GC}) are
shown in Fig. 2 (here $\varepsilon$ is not infinitesimal for
visualization) . To simplify calculations we rotate the contour
$\mathcal{C}$ by a positive angle $\varphi$, and the new contour
$\mathcal{C}'=\mathcal{E}_0e^{i\varphi}t$ is depicted by the dotted
line.

The integral (\ref{GC}) is calculated numerically along the
contour $\mathcal{C}'$. To explain details of integration we
subdivide the path of integration conventionally into two parts:
$\mathcal{C}'_1\ (t<0)$ and $\mathcal{C}'_2\ (t>0)$. We also
introduce for brevity two variables:
$p=\sqrt{\mathcal{E}_0-\mathcal{E}}$ and
$k=\sqrt{\mathcal{E}_0+\mathcal{E}}$.
Note that $\arg(p)>0$ for $t<0$, whereas
$\mathcal{C}_{\nu\,\lambda}^{(-)}(p)$ and $\mathcal{G}_{\nu\,
\nu'}^{\lambda(-)}(p)$ are defined at $-\pi<\arg(p)<0$. To define
$\mathcal{G}_{\nu\, \nu'}^{\lambda(-)}(p)$ on $\mathcal{C}'_1$ we
use the analytic continuation \cite{Broad}
 $\mathcal{C}_{\nu\,\lambda}^{(-)}(p) =
\mathcal{C}_{\nu\,\lambda}^{(+)}(p) -
2{i}\mathcal{S}_{\nu\,\lambda}(p)$, or
\begin{equation}\label{AG}
\mathcal{G}_{\nu\, \nu'}^{\lambda(-)}(p) = \mathcal{G}_{\nu\,
\nu'}^{\lambda(+)}(p)+\frac{2{i}}{p}\mathcal{S}_{\nu\,\lambda}(p)\,\mathcal{S}_{\nu'\,\lambda}(p).
\end{equation}
At the same time, the contour $\mathcal{C}'_1$ passes in the
domain of definition of the function $\mathcal{G}^{\ell\,(-)}_{n\,
n'}(k)$.

Reasoning by analogy allows us to integrate along the contour
$\mathcal{C}'_2$. Finally we obtain in the limit $\varepsilon\to
0$
\begin{equation}\label{I10}
 \begin{array}{c}
 G_{n\,n';\;\nu\,\nu'}^{\ell\,\lambda\,(-)}(E)=
  \frac{\displaystyle \mathcal{E}_0e^{{i}\varphi}}{\displaystyle 2\pi {i}}\left\{ \int
  \limits _{-\infty}^{0}dt\,\mathcal{G}^{\ell\,(-)}_{n\,
n'}(k) \left[\mathcal{G}^{\lambda (+)}_{\nu\, \nu'}(p)
+\frac{2{i}}{p}
\mathcal{S}_{\nu\,\lambda}(p)\,\mathcal{S}_{\nu'\,\lambda}(p)\right]\right.+\\[3mm]
\qquad \qquad \qquad \qquad \qquad  + \left.\int
  \limits _{0}^{\infty}dt\left[\mathcal{G}_{n,
n'}^{\ell(+)}(k)+ \frac{2{
i}}{k}\mathcal{S}_{n\,\ell}(k)\,\mathcal{S}_{n'\,\ell}(k)\right]
\mathcal{G}^{\lambda (-)}_{\nu\, \nu'}(p)\right\}.\\
 \end{array}
\end{equation}
Calculating the second term in (\ref{AG}) we accept in accordance
with \cite{Broad}  that
\begin{equation}\label{G2}
\left|\Gamma(\ell+1+{i}t) \right|^2 \rightarrow
\Gamma(\ell+1+{i}t)\Gamma(\ell+1-{i}t),
\end{equation}
as well as calculating $\xi^{-2{i}t}$ we choose the minimal meaning
of the argument $\xi$ at the range $-\pi<\arg(\xi)\le \pi$.

\section{Results and discussion}

To illustrate the efficiency of the presented numerical scheme,
calculations of the triple differential cross section (TDCS) for
$(e, 2e)$ single ionization and five-fold differential cross
section (5DCS) for $(e, 3e)$ double-ionization reactions on the
helium atom in a singlet state were performed. If a fast
projectile electron of energy about several kiloelectronvolts
transfers to the atom relatively small amounts of energy and
momentum, the respective four-body problem can be sufficiently
simplified by examining only the first Born approximation in the
interaction of the projectile electron with the atom. The
calculations were performed in the limit $m_3=\to\infty$; that is,
${\bf x}_1=\sqrt 2{\bf r}_2$ and ${\bf y}_1=\sqrt 2{\bf r}_1$
(atomic units are used $m_e=e=\hbar=1$).

The triple differential cross section (TDCS) of He$(e, 2e)$He$^+$
reaction when the residual He$^+$ ion remains in an exited state
is written in the form
\begin{equation}
\sigma^{(3)}_{n_0}\equiv\frac{d^3 \sigma_{n_0}}{d\Omega_s
dE_1\,d\Omega_1}=\sum \limits _{\ell_0\, m}\frac{4 p_s k_1}{p_i
Q^4} \left|\left<\Psi^{(-)}_{n_0 \ell_0 m} \right|\exp({i}{\bf
Q}{\bf r}_1)+\exp({i}{\bf Q}{\bf r}_2)-2\left| \Psi_0 \right>
\right|^2,\label{TCS}
\end{equation}
where $(E_i,{\bf p}_i)$, $(E_s,{\bf p}_s)$, and $(E_1,{\bf k}_1)$
are the energy and momentum of, respectively, the incident (fast),
the scattered (fast), and the ejected (slow) electron; ${\bf
Q}={\bf p}_i-{\bf p}_s$ is the transferred momentum.

The five-fold differential cross section (5DCS) of
He$(e,3e)$He$^{++}$ reaction is given by
\begin{equation}
\sigma^{(5)}\equiv\frac{d^5 \sigma}{d\Omega_s dE_1\,d\Omega_1
dE_2\,d\Omega_2}=
\frac{p_s k_1 k_2}{2p_i Q^4} \left|\left<\Psi^{(-)}({\bf k}_1,{\bf
k}_2) \right|\exp({i}{\bf Q}{\bf r}_1)+\exp({i}{\bf Q}{\bf
r}_2)-2\left| \Psi_0 \right> \right|^2,
\label{FCS}
\end{equation}
where  $(E_1, {\bf k}_1)$ and $(E_1, {\bf k}_2)$ are the energies
and momenta of both ejected slow electrons.

The final-state wave functions $\Psi^{(-)}$ (\ref{FWF}) and
$\Psi^{(-)}_{n_0 \ell_0 m}$ (\ref{FWFp}) are obtained using the
method described in the previous Section. We have restricted
ourselves to the maximum value of the total orbital angular
momentum, $L_{max}=2$, and  $\ell,\, \lambda \le 3$. The number of
basis functions (\ref{BF}) for each Jacobi coordinate $x$ and $y$
is taken as $N=21$. The basis parameter $u$ (\ref{BF}) affects the
rate of convergence of the numerical results, and its optimum
value turned out to be $u=0.3$. We also use the following
parameters of the separation function (\ref{zeta}): $\nu=2.5$;
$x_0=0.9$; $y_0=6$.

The helium ground-state wave function $\Psi_0$ is obtained as a
result of diagonalization of the matrix (\ref{H3}) which was
calculated in the basis (\ref{BBF}). The orthonormal functions
\begin{equation}
\varphi_{n}^{\ell}(r)=\left[\frac{(n+1)_{(2\ell+2)}}{2u}\right]^{-1/2}
(2ur)^{\ell+1}e^{-ur}L_{n}^{2\ell+2}(2ur) \label{BFG}
\end{equation}
are used here instead of functions (\ref{BF}). Here we put
$\ell_{max}=3$ and $n_{max}=\nu_{max}=15$. Choosing the basis
parameter $u_0=1.193$ yields the value $E_0=-2.903256$ for the
ground-state energy.

The angular distributions of the slow electron in the case of
$(e,2e)$ ionization-excitation reaction are presented in Figs.
3~-~5, $\theta_1$ being its in-plane emission angle with respect
to the vector ${\bf p}_i$. The displayed experimental data, which
correspond to the general kinematic conditions $n_0=2$, $E_s=5500$
eV, and three particular cases $E_1=5$~eV and $\theta_s=0.35^{o}$
(Fig. 3), $E_1=10$ eV and $\theta_s=0.32^{o}$ (Fig.4), and
$E_1=75$ eV and $\theta_s=1^{o}$ (Fig. 5), where $\theta_s$ is the
scattering angle, are borrowed from~\cite{DLD}.

The results for $\sigma^{(5)}$ are presented in Figs. 6 and 7. The
in-plane angle $\theta_1$ of one of the two slow electrons is
fixed, while the in-plane angle $\theta_2$ of the other slow
electron varies. The energy of the scattered electron $E_s=5500$
eV and its in-plane angle $\theta_s=0.45^{o}$ is also fixed in all
experiments. The energies of the slow electrons are $E_1=E_2=10$
eV (Fig.~6) and $E_1=E_2=4$ eV (Fig.~7). One can see that our
results quite satisfactorily agree with the experimental
distributions both in shape and in absolute value. This agreement
in absolute scale favorably distinguishes our calculations from
that obtained earlier in \cite{KBLDT,KNP} by the method of
pseudostates. Considerable scaling factors were needed there to
compare theory and experiment.

Let us discuss this success of our treatment in more detail. The
exact final state wave function $|\Psi^{(-)}({\bf k}_1,{\bf
k}_2)>$ must be normalized
$$
<\Psi^{(-)}({\bf k}'_1,{\bf k}'_2)|\Psi^{(-)}({\bf k}_1,{\bf
k}_2)>=\delta({\bf k}_1-{\bf k}'_1)\delta({\bf k}_2-{\bf k}'_2)+
\delta({\bf k}_1-{\bf k}'_2)\delta({\bf k}_2-{\bf k}'_1)
$$
The symmetrized sum of plane waves obviously possesses this
property, as well as do Coulomb waves.  Recent calculations with a
3C function \cite{JM,AMC}
\begin{equation}
\Psi^{(-)}_{3C} ({\bf k}_1,{\bf k}_2; {\bf r}_1,{\bf r}_2) =
\frac{(1+gP_{12})}{\sqrt{2}} \ e^{-i{\bf k}_{12}{\bf r}_{12}}
\varphi_1^{-}({\bf k}_1, {\bf r}_1) \varphi_2^{-}({\bf k}_2, {\bf
r}_2) \varphi_{12}^{-}({\bf k}_{12}, {\bf r}_{12}) \label{bbk}
\end{equation}
and some variational ground-state helium functions have
demonstrated striking agreement with the experiment, although the
published evidences that this function is normalized are unknown
to us.

It is also obvious that the function $\Psi^{(-)}$ being decomposed
into a {\it finite} sum of square-integrable functions can never
be normalized to $\delta$ function. Our numerical scheme allows to
conclude that the asymptotic bound of the function $\Psi^{(-)}$ is
a product of two Coulomb functions in the discrete representation.
Presumably, it is this property which provides much better
normalization conditions for the continuum wave function than in
the case of the pseudostates' approach.

As a summary, we formulate two main conclusions
\begin{itemize}

\item  The proposed numerical scheme and calculations demonstrate
the importance of accounting for the whole two-body continuum
spectrum. The method of pseudostates which replaces the continuum
by a finite number of states with positive energies, seems to face
with serious difficulties as to the magnitude of the calculated
differential cross sections, especially when the resulting
final-state wave function is applied to the calculation of
$(e,3e)$ matrix elements \cite{KBLDT,KNP}.

\item We observe a lack of coincidence between the theory and
experiment in some kinematical cases. This, perhaps, testifies to
a necessity of taking into account the correct behavior of the
function $\Psi^{(-)}$ in a "true" \ three-body asymptotic region.
The simple 3C model (\ref{bbk}) partially accounts for such a
behavior.

\end{itemize}

Finally we can conclude that the presented method based on the
$J$-matrix approach allows to formulate the effective numerical
scheme for applications in atomic physics.

\appendix*{}
\section{}

The Hamiltonian
\begin{equation}\label{RH}
    H_{\ell}=-\frac{d^2}{dr^2}+\ell(\ell+1)/r^2+{2Z}/r
\end{equation}
has the following wave functions of the bound spectrum ($Z<0$)
\begin{equation}\label{ABndf} \varphi^{(j)}_{\ell}(r) = (2\kappa_j
\, r)^{\ell+1}e^{-\kappa_j\,r}
\frac{\sqrt{-Z\,(j+1)_{2\ell+1}}}{(j+\ell+1)(2\ell+1)!}\; {_1
F_1}(-j; \; 2\ell+2; \; 2\kappa_j \, r),
\end{equation}
and continuous spectrum
\begin{equation}
\varphi_{\ell}(k, \, r)= \frac{1}{2}(2kr)^{\ell+1}\,e^{-\pi
t/2}\,e^{{i}k r}\frac{\left|\Gamma(\ell+1+{i}t) \right|}{(2\ell+1)!}
\, {_1 F_1}(\ell+1+{i}t;\; 2\ell+2; \; -2{i}k r). \label{ARegSol}
\end{equation}
They can be expanded in the Laguerre basis functions \cite{JA}
\begin{equation}
\phi_{n}^{\ell}(r)=\left[(n+1)_{(2\ell+1)}
\right]^{-1/2}(2ur)^{\ell+1}e^{-ur}L_{n}^{2\ell+1}(2ur) \label{LBF}
\end{equation}
and the expansion coefficients $\mathcal{S}_{n\,\ell}^{(j)}$ and
$\mathcal{S}_{n\,\ell}(k)$ are given by \cite{JA,Broad}
\begin{equation}
 \begin{array}{l}
\mathcal{S}_{n
\ell}^{(j)}=(-1)^n\,\left(\frac{4u\kappa_j}{(u+\kappa_j)^2}\right)^{\ell+1}\,\left(
\frac{u-\kappa_j}{u+\kappa_j}\right) ^{n+j}\,
\frac{\sqrt{-Z\,(j+1)_{2\ell+1}\,(n+1)_{2\ell+1}}}
{(j+\ell+1)(2\ell+1)!} \,
 \times \\[3mm]
\qquad \quad \qquad \qquad \qquad \qquad \qquad \qquad \qquad \times
{_2F _1}\left(-n, -j; \; 2\ell+2;\; 1-
\left( \frac{u+\kappa_j}{u-\kappa_j}\right)^2 \right)\\
 \end{array}
  \label{ASB}
\end{equation}
and
\begin{equation}
 \begin{array}{c}
\mathcal{S}_{n \ell}(k)= \frac{1}{2}\,\sqrt{(n+1)_{2\ell+1}}\,(2\sin
\zeta)^{\ell+1}\,e^{-\pi t/2}\,\xi^{-{i}t}\frac{\displaystyle
\left|\Gamma(\ell+1+{i}t) \right|}{\displaystyle (2\ell+1)!}\,
\times \qquad \qquad \qquad \qquad\\[3mm]
\qquad \qquad \qquad \qquad \qquad \times (-\xi)^n\, {_2
F_1}\left(-n, \,
\ell+1+{i}t; \; 2\ell+2; \; 1-\xi^{-2}\right).\\
 \end{array}
 \label{ASsol}
\end{equation}
Here $\kappa_j=-{Z}/{(j+\ell+1)}$, $t={Z}/{k}$, and
$\xi=e^{{i}\zeta}=({{i}u-k })/({{i}u+k})$.

It is proved \cite{JA} that the functions $\mathcal{S}_{n
\ell}^{(j)}$, $\mathcal{S}_{n\,\ell}$ are the regular solutions of
the infinite trinomial recurrence which is a discrete analog of
the Schr\"odinger equation
\begin{equation}\label{ATRR}
    \mathcal{J}_{n,\, n-1}^{\ell}(k)d_{n-1}+\mathcal{J}_{n,\, n}^{\ell}(k)d_n+
    \mathcal{J}_{n,\,
    n+1}^{\ell}(k)d_{n+1}=0 \qquad (n=1,2,\ldots)
\end{equation}
with the initial condition
\begin{equation}\label{BC}
\mathcal{J}_{0,\, 0}^{\ell}(k)d_{0}+\mathcal{J}_{0,\,
1}^{\ell}(k)d_1=0.
\end{equation}
In (\ref{ATRR}) $\mathcal{J}_{n,\, n'}^{\ell}(k)$ are the elements
of three-diagonal matrix of the operator $(H_{\ell}-k^2)$ (i.e.
$J$-matrix) calculated with the basis functions $\phi_{n}^{\ell}$
\begin{equation}
 \begin{array}{c}
\mathcal{J}_{n\,n}^{\ell}(k)=\frac{u^2-k^2}{u}(n+\ell+1)+2Z,\\[3mm]
\mathcal{J}_{n\,n-1}^{\ell}(k)= \frac{
u^2+k^2}{2u}\sqrt{n(n+2\ell+1)},\\[3mm]
\mathcal{J}_{n\,n+1}^{\ell}(k)= \frac{
u^2+k^2}{2u}\sqrt{(n+1)(n+2\ell+2)}.\\
 \end{array}
 \label{AJMT}
\end{equation}

The coefficient functions $\mathcal{S}_{n \ell}^{(j)}$,
$\mathcal{S}_{n\,\ell}$ satisfy the normalization conditions
\begin{equation}
 \begin{array}{c}
  \sum \limits _{n,\,n'=0}^{\infty}\mathcal{S}_{n\ell}^{(j)}\,Q_{n, \,
  n'}^{\ell}\,
  \mathcal{S}_{n'\ell}^{(j')}=\delta_{j, \, j'},\\[4mm]
\frac{2}{\pi}  \sum \limits
_{n,\,n'=0}^{\infty}\mathcal{S}_{n\ell}(k)\,Q_{n, \, n'}^{\ell}\,
  \mathcal{S}_{n'\ell}(k')=\delta(k-k'),\\
 \end{array}
\label{Norm}
\end{equation}
where $Q_{n\,n'}^{\ell}$ are the elements of three-diagonal matrix
and are given by the overlapping integrals
\begin{equation}
Q_{n\,n'}^{\ell}=\int_0^{\infty}
\phi_n^{\ell}(r)\phi_{n'}^{\ell}(r) dr,\label{QI}
\end{equation}
\begin{equation}
 \begin{array}{c}
Q_{n\,n-1}^{\ell}= -\frac{\displaystyle 1}{\displaystyle
2u}\sqrt{n(n+2\ell+1)}, \quad Q_{n\,n+1}^{\ell}=
-\frac{\displaystyle 1}{\displaystyle
2u}\sqrt{(n+1)(n+2\ell+2)},\\[3mm]
Q_{n\,n}^{\ell}= \frac{\displaystyle 1}{\displaystyle
u}(n+\ell+1).\\
 \end{array}
  \label{QT}
\end{equation}
The corresponding completeness condition takes the form
\begin{equation}
\frac{2}{\pi} \int \limits _0 ^{\infty} dk\, Q_{n, \, n''}^{\ell}
\mathcal{S}_{n'' \ell}(k)\mathcal{S}_{n'\ell}(k)+\sum_{j=0}^{\infty}
Q_{n, \, n''}^{\ell} \mathcal{S}_{n'' \ell}^{(j)}\,\mathcal{S}_{n'
\ell}^{(j)} =\delta_{n\,n'}.\label{Complete}
\end{equation}

The irregular solution of Eq.~(\ref{ATRR}) can be written in the
form \cite{JA,Broad}
\begin{equation}
 \begin{array}{c}
\mathcal{C}_{n\,\ell}^{(\pm)}(k)= -\sqrt{n!(n + 2\ell+1)!}
\,\frac{\displaystyle e^{\pi t/2}\xi^{{i}t}}{\displaystyle (2\sin
\zeta)^{\ell}}\,\frac{\displaystyle \Gamma(\ell+1\pm{i}t)}
{\displaystyle \left|\Gamma(\ell+1\pm{i}t)\right|}\times \qquad \qquad \qquad \\[3mm]
\qquad \qquad\qquad \qquad \times \frac{\displaystyle
(-\xi)^{\pm(n+1)}}{\displaystyle \Gamma(n+\ell+2\pm{i}t)} \,
{_2F_1}\left(-\ell\pm{i}t,
\, n+1; \; n+\ell+2\pm{i}t; \; \xi^{\pm 2}\right). \\
 \end{array}
 \label{ASCp}
\end{equation}
The function $\mathcal{C}_{n\,\ell}^{(+)}(k)$
$\left(\mathcal{C}_{n\,\ell}^{(-)}(k)\right)$ is determined at
${\rm Im}({k})>0$ [${\rm Im}({k})<0$] of the complex plane $k$.
The analytical continuation can be done with the help of the
following relation \cite{Broad}:
\begin{equation}\label{AC}
\mathcal{C}_{n\,\ell}^{(+)}(k) = \mathcal{C}_{n\,\ell}^{(-)}(k) +
2{i}\mathcal{S}_{n\,\ell}(k).
\end{equation}

\newpage
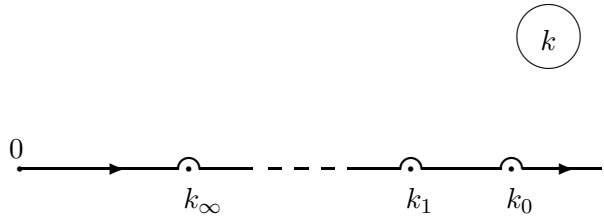
\begin{figure}[ht]
$$
\thicklines
\begin{picture}(260,80) \put(40,20){\circle*{1.5}}
\put(40,20){\line(1,0){60}}\put(60,20){\vector(1,0){20}}
\put(36,24){$0$}\put(104,20){\oval(8,8)[t]}
\put(104,20){\circle*{2}}\put(102,5){$k_{\infty}$}\put(108,20){\line(1,0){20}}
\multiput(134,20)(10,0){3}%
{\line(1,0){5}}\put(164,20){\line(1,0){20}}\put(188,20){\oval(8,8)[t]}
\put(188,20){\circle*{2}}\put(186,5){$k_{1}$}
\put(192,20){\line(1,0){30}}\put(226,20){\oval(8,8)[t]}
\put(226,20){\circle*{2}}\put(224,5){$k_{0}$}
\put(230,20){\line(1,0){30}}\put(230,20){\vector(1,0){20}}
{\thinlines\put(240,70){\circle{25}}}\put(237,65){$k$}
\end{picture}
$$
\caption{The integration contour of the integral in parentheses in
Eq. (\ref{CE}). The poles of the integrand are shown as closed
circles.}
\end{figure}

\newpage
\begin{figure}[ht]
$$
\thicklines
\begin{picture}(240,160)
\put(0,80){\line(1,0){240}} \put(120,0){\line(0,1){160}}
\put(125,150){$\mbox{\small
Im}\mathcal{E}$}\put(230,67){$\mbox{\small Re}\mathcal{E}$}
\put(160,80){\vector(1,0){40}} \put(195,67){$\mathcal{C}$}
\put(100,100){\oval(6,6)[l]}\put(100,103){\line(1,0){139}}
\put(100,97){\line(1,0){139}}\put(75,107){${
-\mathcal{E}_0+{i}\varepsilon}$}\put(97,100){\circle*{3}}
\put(80,100){\circle*{3}}\put(50,100){\circle*{3}}\put(15,100){\circle*{3}}
\put(140,60){\oval(6,6)[r]}\put(0,63){\line(1,0){140}}
\put(0,57){\line(1,0){140}}\put(130,40){${
\mathcal{E}_0-{i}\varepsilon}$}\put(143,60){\circle{3}}
\put(151,60){\circle{3}}\put(170,60){\circle{3}}\put(200,60){\circle{3}}
\put(100,60){\line(1,1){40}}\multiput(142,102)(15,15){4}%
{\line(1,1){10}}
\multiput(42,2)(15,15){4}%
{\line(1,1){10}} \put(120,80){\vector(1,1){10}}
\put(93,67){$\mathcal{C}'$} \put(134,80){\oval(30,30)[tr]}
\put(134,83){$\varphi$}
\end{picture}
$$
\caption{The path of integration of the convolution integral
(\ref{GC}). The poles which are associated with the discrete
spectrum of the Hamiltonian $h_x$ are depicted as closed circles.
$-\mathcal{E}_0+{i}\varepsilon$ denotes the initial point of the
corresponding unitary branch cut. Open circles and
$\mathcal{E}_0-{i}\varepsilon$ are the same for the Hamiltonian
$h_y$.}
\end{figure}
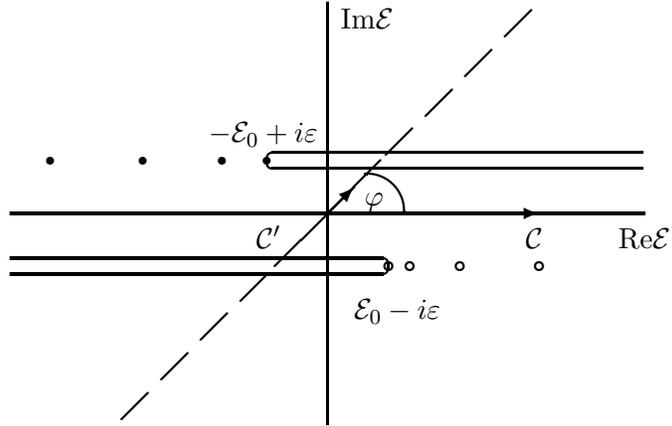

\newpage
\begin{figure}[ht]
\centerline{\psfig{figure=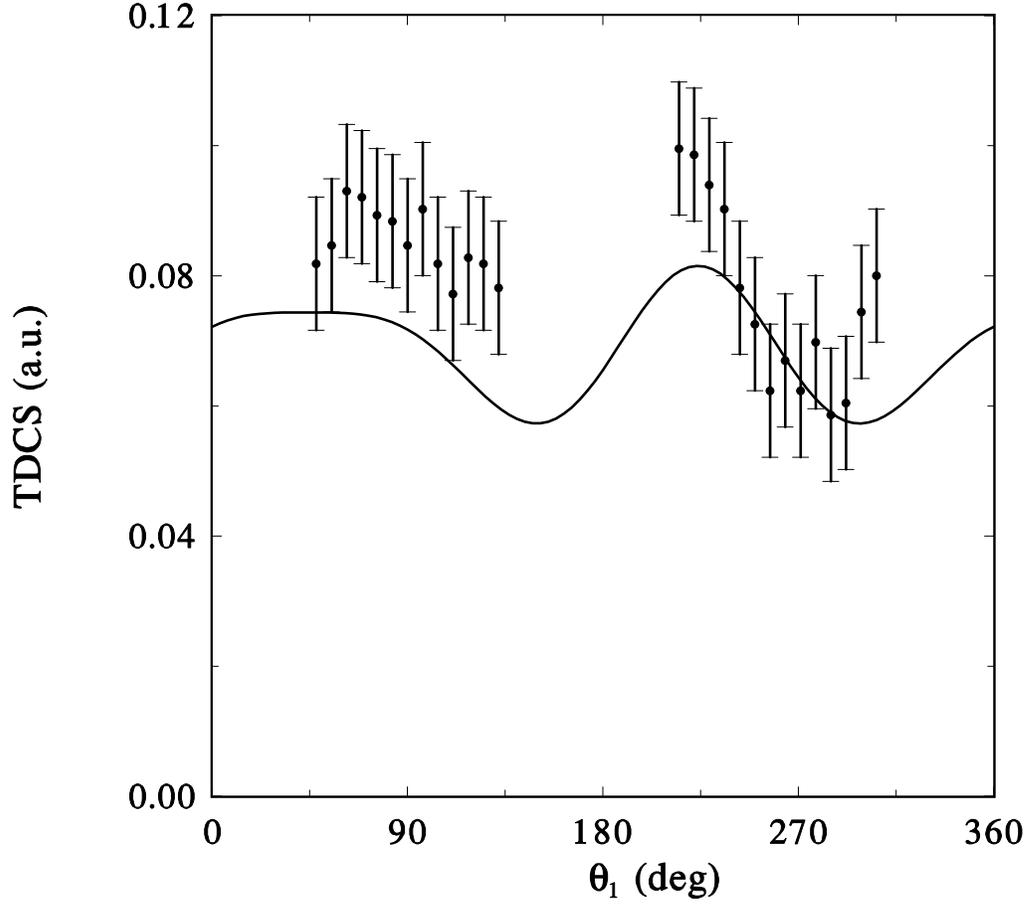,width=1\textwidth}} \caption{TDCS
for the ionization-excitation reaction He$(e,2e){\rm He}^+$ when the
helium ion is left in the state $n_0=2$. The fast scattered electron
energy is $E_s=5500$ eV, while $E_1=5$ eV for the slow ejected
electron. The scattering angle of the fast incident electron is
fixed $\theta_s=0.35^o$, and the angle of the ejected electron
$\theta_1$ varies relative to the incident electron direction. All
electron velocities are disposed on the same plane. The experimental
data are borrowed from \cite{DLD}.}
\end{figure}

\newpage
\begin{figure}[ht]
\centerline{\psfig{figure=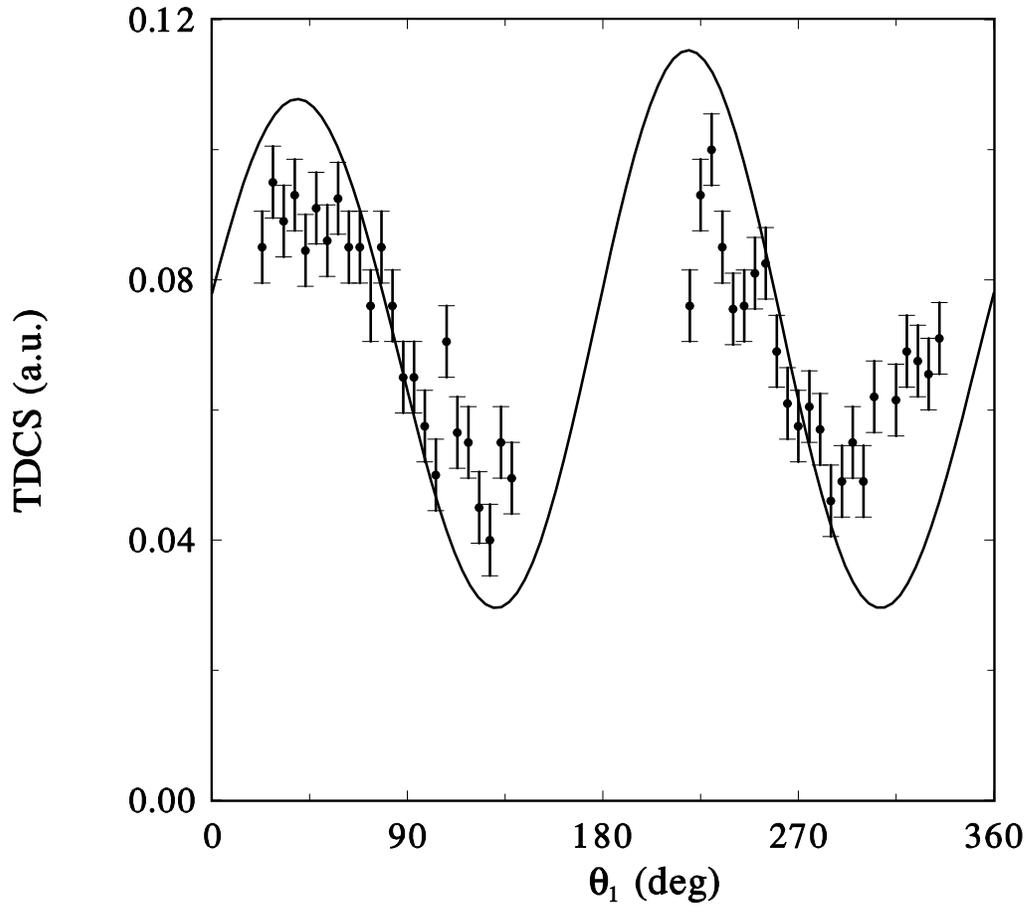,width=1\textwidth}} \caption{ The
same as in Fig.~3, except $E_1=10$ eV and $\theta_s=0.32^o$. }
\end{figure}

\newpage
\begin{figure}[ht]
\centerline{\psfig{figure=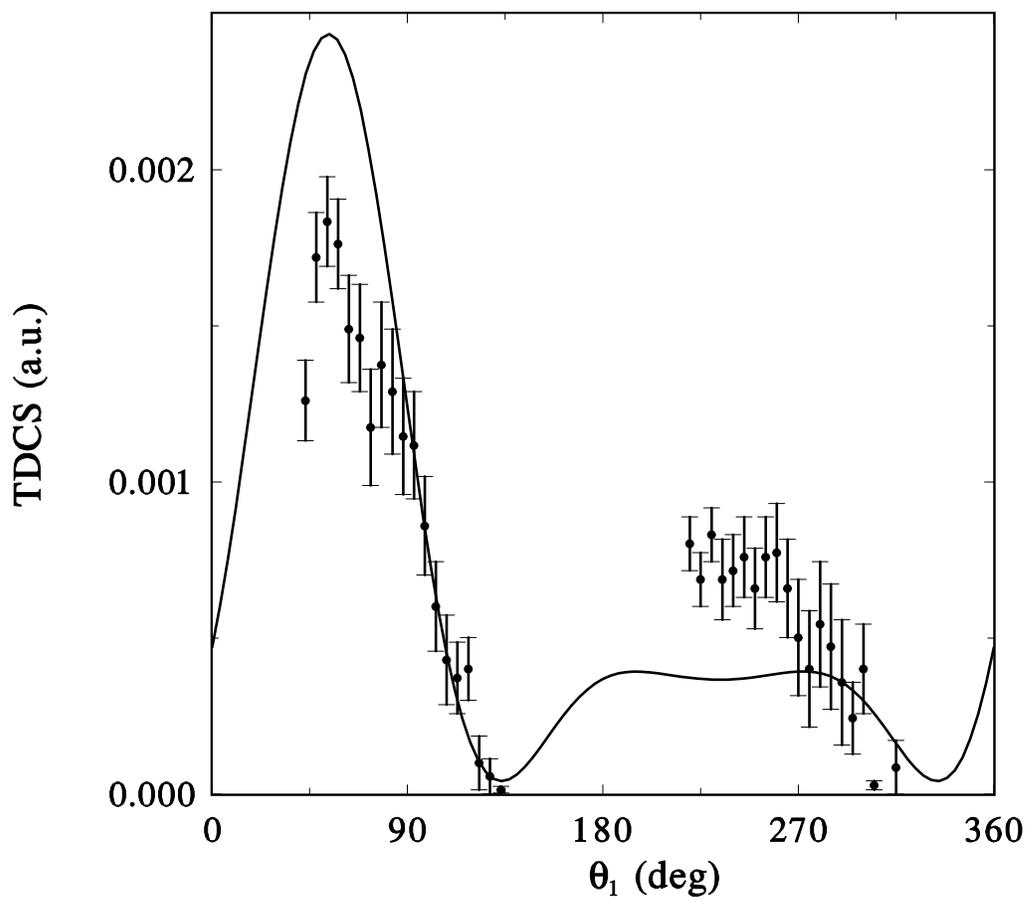,width=1\textwidth}} \caption{The
same as in Fig.~3, except $E_1=75$ eV and $\theta_s=1^o$.}
\end{figure}

\newpage
\begin{figure*}[ht]
\centerline{\psfig{figure=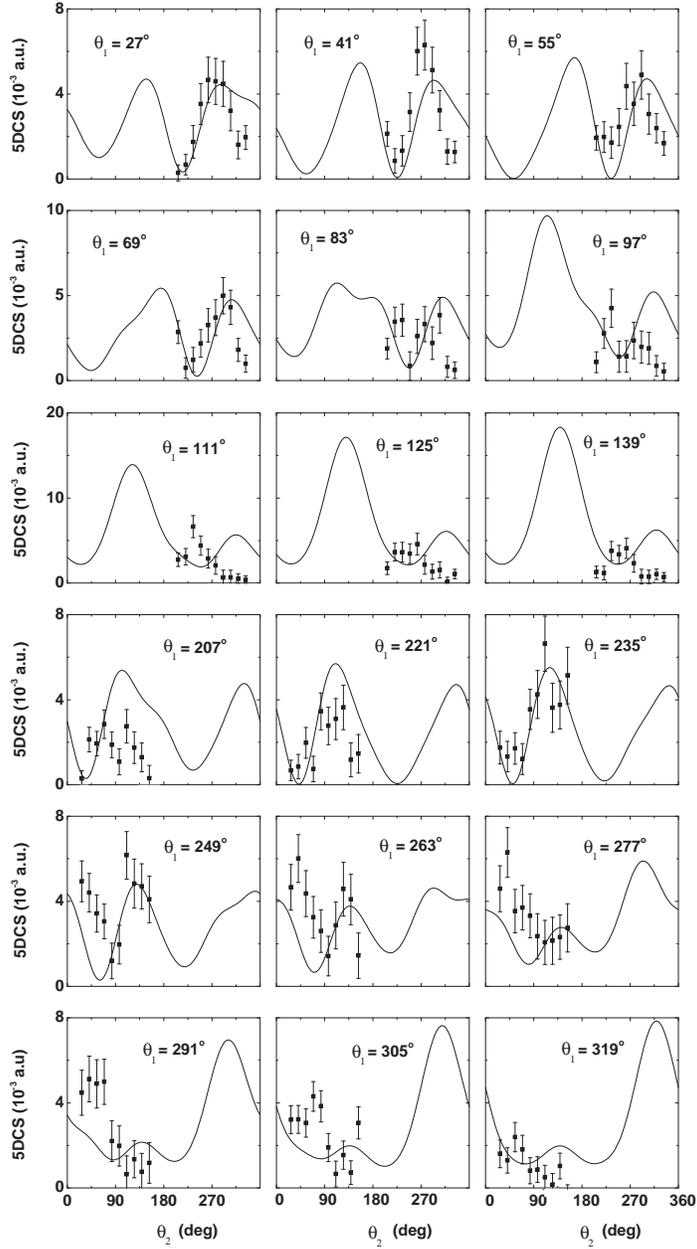,width=0.625\textwidth}}
\caption{The fully resolved five-fold differential cross section
{5DCS} of the electron-impact double ionization reaction
He$(e,3e){\rm He}^{++}$. The energy of the scattered electron is
$E_s=5500$ eV, and the energies of the slow ejected electrons are
$E_1=E_2=10$ eV. The scattering angle of the fast incident electron
is fixed $\theta_s=0.45^o$, and the angles of the ejected electrons
are $\theta_1$ and $\theta_2$, where one angle is fixed while the
other angle varies relative to the incident electron direction. All
electron velocities are disposed on the same plane. The absolute
measurements are borrowed from \cite{C4FS}.}
\end{figure*}

\newpage
\begin{figure*}[ht]
\centerline{\psfig{figure=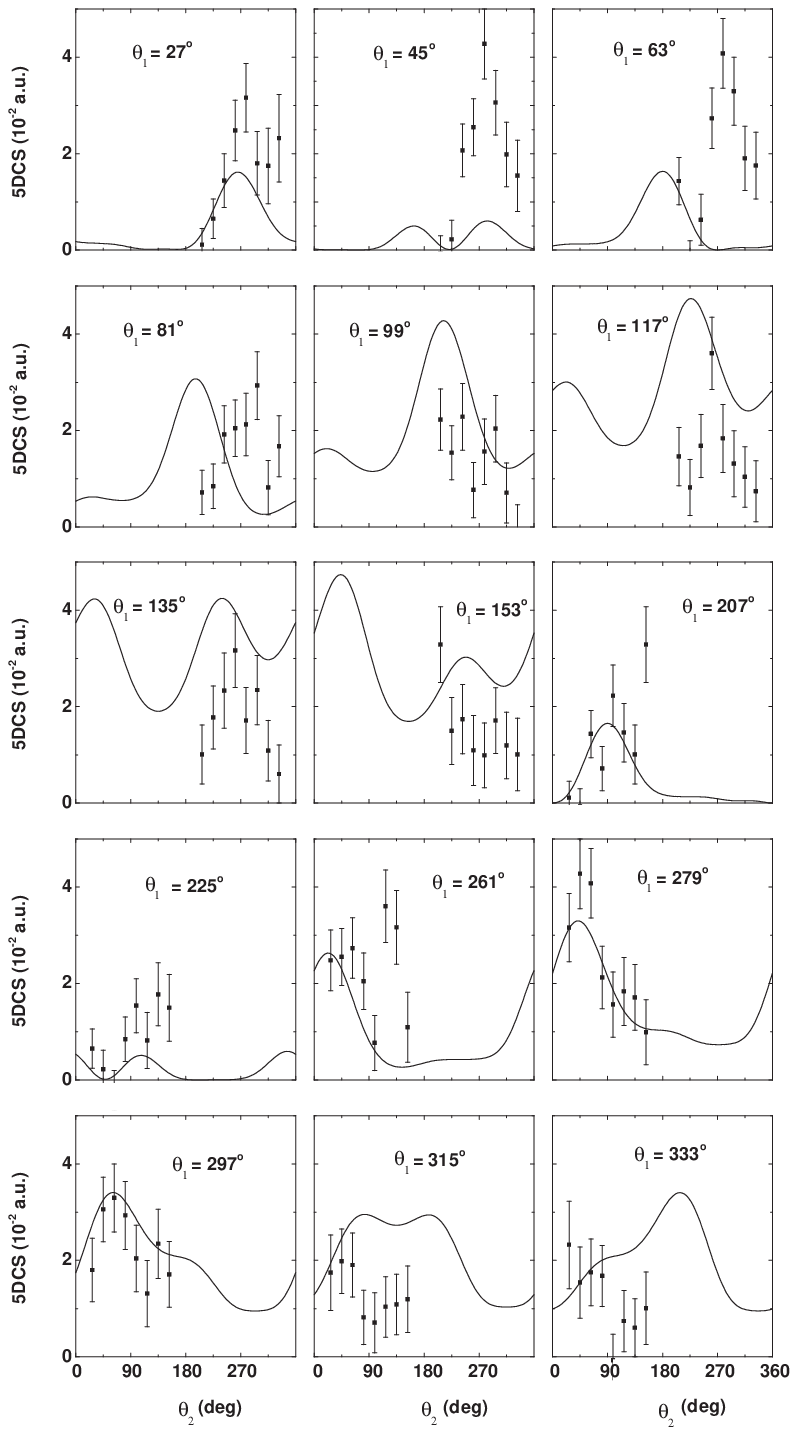,width=0.75\textwidth}}
\caption{The same as in Fig.~6, except the ejection energies are
lowered to $E_1=E_2=4$ eV.}
\end{figure*}


\newpage
\begin{thebibliography}{99}


\bibitem{JA}
 E.~J.~Heller and H.~A.~Yamani, Phys. Rev. A {\bf 9}, 1201 (1974);
 H.~A.~Yamani and L.~Fishman, J. Math. Phys. {\bf 16}, 410 (1975);
 J.~T.~Broad and W.~P.~Reinhardt, Phys. Rev. A {\bf 14}, 2159 (1976).

\bibitem{FO}
 G.~F.~Fillipov and I.~P.~Okhrimenko, Sov.J.Nucl.Phys {\bf 32}, 480 (1980);
 G.~F.~Fillipov, Sov.J.Nucl.Phys {\bf 33}, 488 (1981).

\bibitem{SN}
 Yu.~F.~Smirnov and Yu.~I.~Nechaev, Sov.J.Nucl.Phys  {\bf 35}, 808 (1982).

\bibitem{RSJ}
 J.~Revai, M.~Sotona, and J.~Zofka, J. Phys. G {\bf 11}, 745 (1985).

\bibitem{Rmtx}
 A.~M.~Lane and R.~G.~Thomas, Rev. Mod. Phys. {\bf 30}, 257 (1958).

\bibitem{Papp}
 Z.~Papp, Phys. Rev. C {\bf 55}, 1080 (1997).

\bibitem{BS}
I.~Bray and A.~T.~Stelbovics, Phys. Rev. A {\bf 46}, 6995 (1992).

\bibitem{KS}
 V.~A.~Knyr and L.~Ya.~Stotland,  Sov.J.Nucl.Phys {\bf 55}, 1626 (1992); Phys.At.Nucl. {\bf 56},
 886 (1993); Phys.At.Nucl. {\bf 59}, 575 (1996).

\bibitem{ZKP}
S.~A.~Zaitsev, V.~A.~Knyr, and Yu.~V.~Popov, Phys.At.Nucl. {\bf
69}, 255 (2006).

\bibitem{PHHKY}
 Z.~Papp, C-.~Y.~Hu, Z.~T.~Hlousek, B.~Konya, S.~L.~Yakovlev,
 Phys. Rev. A {\bf 63}, 62721 (2001).

\bibitem{MFB}
 S.~P.~Merkuriev and L.~D.~Faddeev, {\sl Quantum Scattering Theory for Several Particle Systems}
(Kluwer Academic Publishers, Dordrecht, 1993).

\bibitem{KBLDT}
 A.~Kheifets, I.~Bray, A.~Lahmam-Bennani, {\it et al.},
  J. Phys. B {\bf 32}, 5047 (1999).

\bibitem{KNP}
V.~A.~Knyr, V.~V.~Nasyrov, and Yu.~V.~Popov, JETP {\bf 92}, 789
(2001); in {\it Correlation and Polarization in Photonic,
Electronic, and Atomic Collisions}, eds. G.F. Hanne, L. Malegat
and H. Schmidt-B\"ocking, AIP Conf.Proc. {\bf 697}, 76 (2003).

\bibitem{JM}
 S.~Jones and D.~H.~Madison, Phys. Rev. Lett. {\bf 91}, 073201
(2003); S. ~Jones, J.~H. ~Macek, and D.~H. ~Madison. Phys. Rev. A
{\bf 70}, 012712 (2004).

\bibitem{AMC}
L.~U. ~Ancarani, T. ~Montagnese, and C. ~Dal Cappello, Phys. Rev.
A {\bf 70}, 012711 (2004).

\bibitem {6C} A.~W.~Malcherek and J.~S.~Briggs, J.~Phys.~B {\bf 30},
4419 (1997); J.~R.~Gotz, M.~Walter, and J.~S.~Briggs, J.~Phys.~B
{\bf 36}, L77 (2003).

\bibitem {C4FS} A.~Lahmam-Bennani, I.~Taouil, A.~Duguet {\it et
al.}, Phys.Rev.A {\bf 59}, 3548 (1999).

\bibitem{KWH}
A.~A.~Kvitsinsky, A.~Wu, and C.-Yu ~Hu,  J.~Phys.~B {\bf 28}, 275
(1995).

\bibitem{Broad}
J.~T.~Broad, Phys. Rev. A, {\bf 31}, 1494 (1985).

\bibitem{Heller}
E.~J.~Heller, Phys. Rev. A {\bf 12}, 1222 (1975).

\bibitem{RS}
 R.~Shakeshaft, Phys. Rev. A, {\bf 70}, 042704 (2004).

\bibitem{DLD}
 C.~Dupr\'e, A.~Lahmam-Bennani, A.~Duguet et al, J. Phys. B, {\bf 25}, 259
(1992).








\end{thebibliography}
\end{document}